\documentclass[a4paper,11pt]{article}
\pdfoutput=1
\usepackage{jinstpub}

\title{Performance of a resistive plate chamber equipped
with a new prototype of amplified front-end electronics}

\collaboration[c]{on behalf of the ALICE Collaboration}

\author[a,b]{Massimiliano Marchisone}
\affiliation[a]{iThemba LABS, National Research Foundation, PO Box 722, Somerset West, South Africa}
\affiliation[b]{University of the Witwatersrand, 1 Jan Smuts Avenue, Braamfontein, Johannesburg, South Africa}

\emailAdd{mmarchis@cern.ch}

\abstract{ALICE is the LHC experiment dedicated to the study of heavy-ion collisions.
At forward rapidity a muon spectrometer detects muons from low mass mesons,
quarkonia, open heavy-flavor hadrons as well as weak bosons.

A muon selection based on transverse momentum is made by a trigger system
composed of 72 resistive plate chambers (RPCs). For the LHC Run 1 and the
ongoing Run 2 the RPCs have been equipped with a non-amplified FEE called ADULT.
However, in view of an increase in luminosity expected for Run 3 (2021--2023)
the possibility to use an amplified FEE has been explored in order to improve 
the counting rate limitation and to prevent the aging of the detector, by
reducing the charge per hit. A prototype of this new electronics (FEERIC) has
been developed and tested first with cosmic rays before equipping one RPC
in the ALICE cavern with it.

In this paper the most important performance indicators - efficiency, dark
current, dark rate, cluster size and total charge - of an
RPC equipped with this new FEE will be reviewed and compared to the others
read out with ADULT, in pp collisions at $\sqrt{s}=5$ and 13 TeV
and in Pb--Pb collisions at $\sqrt{s_{\mathrm{NN}}}=5$ TeV.}

\arxivnumber{1605.08537}
\subheader{XIII Workshop on Resistive Plate Chambers and related detectors,\\Ghent 22--26/02/2016}

\begin{document}
\maketitle

\section{Introduction}
ALICE \cite{ALICE} is a multitasking experiment at the CERN--LHC, mainly designed to study
the hot and dense matter produced in heavy-ion collisions. At forward rapidity ($2.5<y<4$) muons
are detected and tracked with a muon spectrometer \cite{spettrometro}. It consists of a set of absorbers,
a dipole magnet, a muon tracking system and a muon trigger detector \cite{muon_trigger}.

The muon trigger system consists of 72 Resistive Plate Chambers (RPCs) \cite{RPC} arranged in
two stations at a distance of 16 and 17 m from the interaction point. Each
station is made of two detection planes with 18 RPCs each, perpendicular to the beam line.
The chambers are 2 mm single gap RPC, with low resistivity ($\sim$10$^9$ $\Omega\cdot$cm) \cite{bakelite}
bakelite electrodes and are operated in the so-called ``maxi-avalanche'' mode \cite{maxi-avalanche},
with a gas mixture consisting of 89.7\% C$_2$H$_2$F$_4$, 10\% C$_4$H$_{10}$ and 0.3\% SF$_6$.
The gas relative humidity is kept at 37\% to prevent alterations
in the bakelite resistivity \cite{humidity}. The operating high voltages are optimized for each RPC and range
from 10.0 to 10.4 kV \cite{RPC commissioning}.
The signal is picked up inductively on both sides of the detector by means of 
orthogonal copper strips (1, 2 and 4 cm), in the horizontal (bending plane)
and vertical (non-bending plane) directions.
With the present FEE (ADULT ASIC \cite{ADULT}) the signal is discriminated without
pre-amplification, with thresholds set at 7 mV and the mean charge per hit produced
in the gas gap is of the order of 100 pC \cite{RPC_run1}.

In those conditions, the instantaneous counting rate limit is 50 hits/s/cm$^2$,
including some safety margins. Furthermore, the safe operation of the detectors cannot be
guaranteed for a cumulative charge larger than 50 mC/cm$^2$ (i.e. 500 Mhits/cm$^2$ in maxi-avalanche)
\cite{maxi-avalanche}. These two limitations are not compatible with the expected conditions
after the second LHC long shutdown (2019--2020). Indeed, peaks of counting rates up to 125 hits/s/cm$^2$
and a cumulative charge of 100 mC/cm$^2$ should be reached by the most exposed RPCs in central Pb--Pb 
collisions.

A possible solution for overcoming these two problems is to operate the RPCs at lower gain,
like in ATLAS \cite{ATLAS1,ATLAS2} and in CMS \cite{CMS1,CMS2}.
This requires an upgrade of electronics with amplification to reduce the charge per hit.
The proposed solution is to replace the present front-end boards with new ones based on the
FEERIC ASIC (Front-End Electronics Rapid Integrated Circuit) developed at Clermont-Ferrand (France)
\cite{FEERIC}. Unlike ADULT, FEERIC performs
amplification of the analog signals from the RPC before discrimination. This should
allow one to reduce by a factor 3--5 the charge produced in the gas, hence limiting the aging effects.

\section{First results with cosmic rays}
Preliminary measurements have been performed with FEERIC board prototypes for strips of 2 cm at 
the beginning of 2015, using the test bench available in Torino (Italy) \cite{test_bench}.
The RPC under test has been filled with the same gas mixture and relative humidity adopted in the
muon trigger.

\begin{figure}[htbp]
 \centering
 \includegraphics[width=0.496\textwidth]{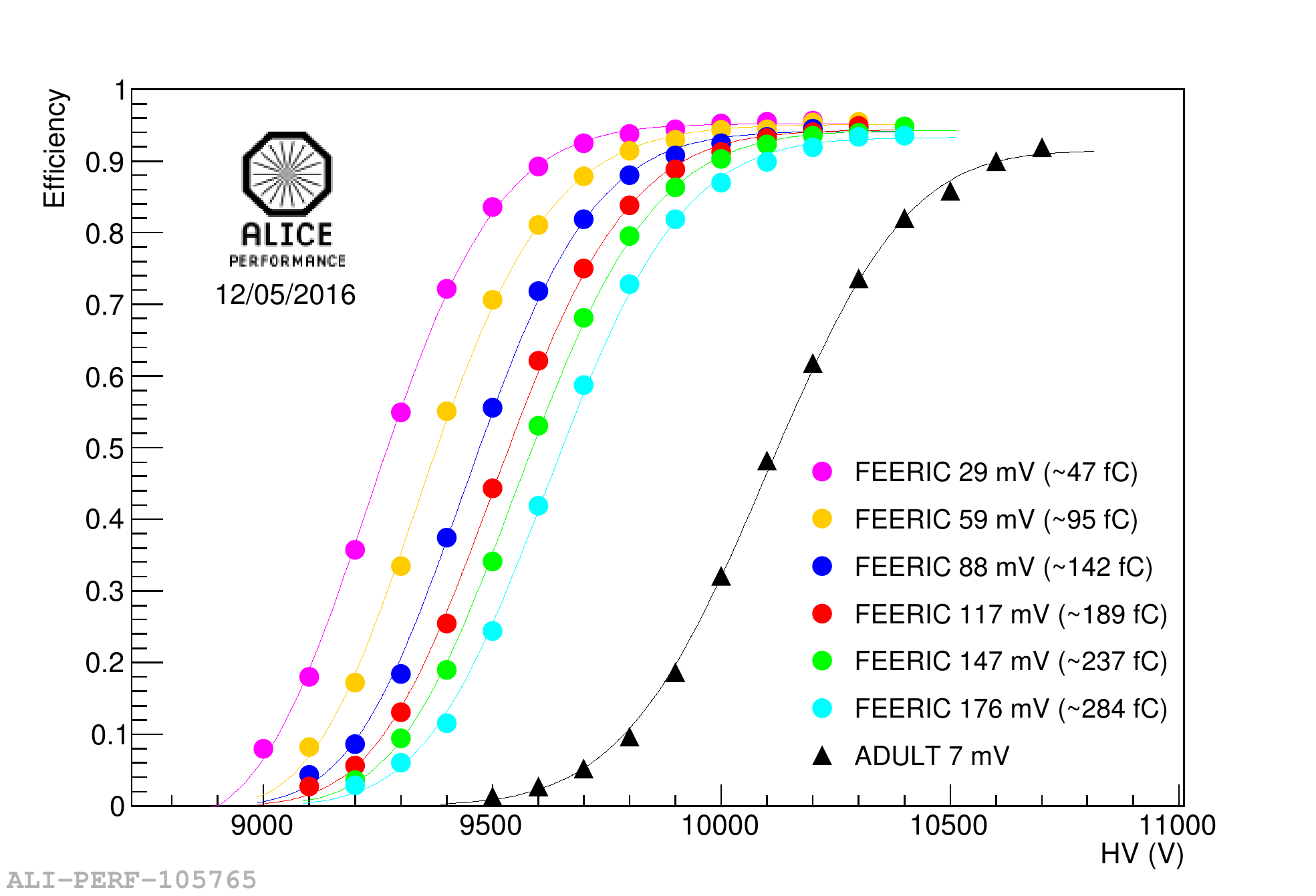}
 \includegraphics[width=0.496\textwidth]{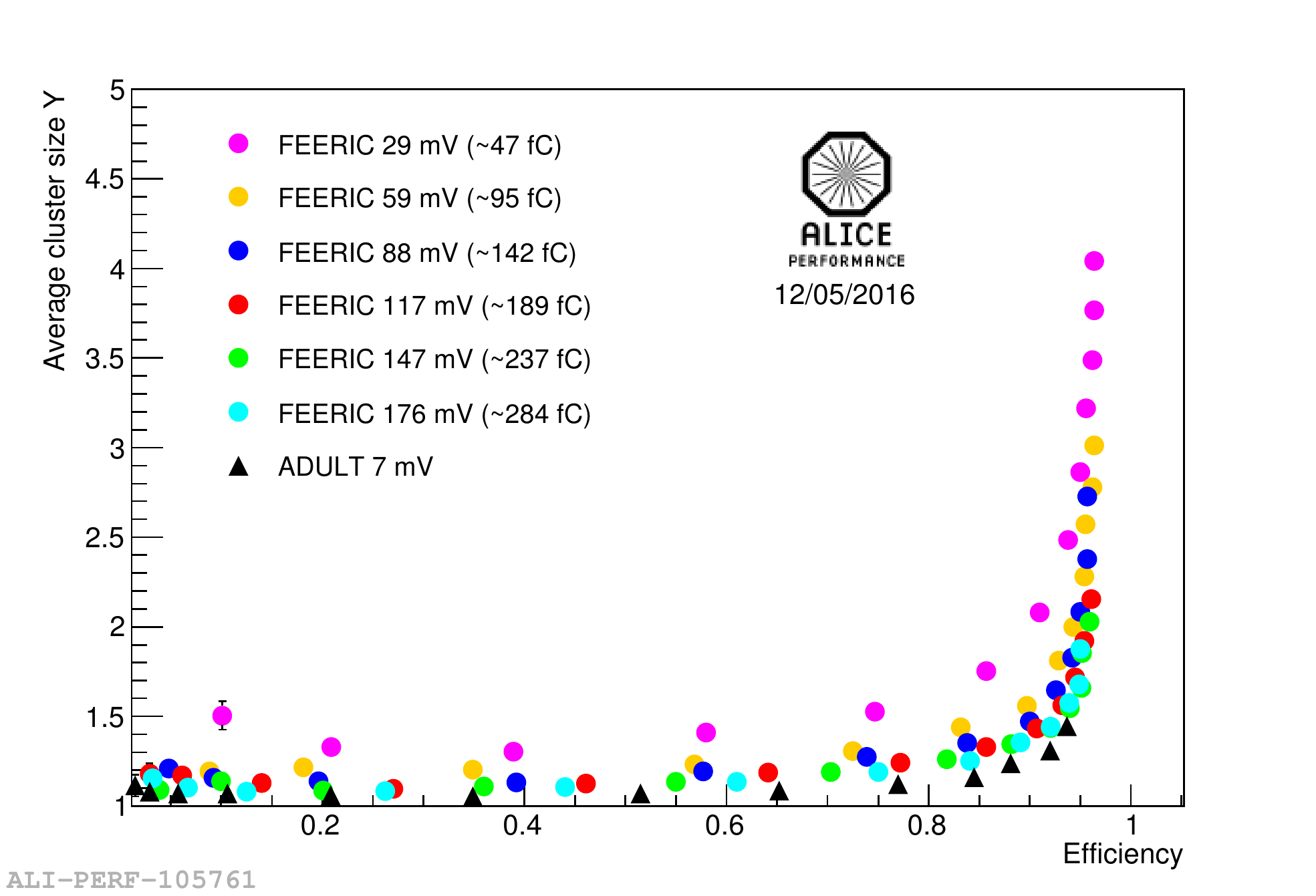}
 \caption{Efficiency curve (left) and cluster size measurements for strips of 2 cm (right) from
 cosmic data taking. Colors refer to different discrimination thresholds.}
 \label{fig: cosmic_results}
\end{figure}

Efficiency curves with FEERIC for various discrimination thresholds are compared in the left
figure \ref{fig: cosmic_results}, to the one obtained with ADULT with 7 mV threshold.
The rise to the efficiency plateau is
found to be quite sharp and a clear shift towards lower high voltage working points can be
observed with FEERIC. The HV shift depends strongly on the FEERIC discrimination threshold, which
is limited by the environmental noise conditions, however the HV shift is not less than 600--700 V.

The cluster size as a function of the RPC efficiency is shown in
figure \ref{fig: cosmic_results} (right). In general the average value measured at the working point with
FEERIC is slightly higher than what is measured with ADULT. This can be explained by considering the different
signal amplitude distributions in the two working modes.

\section{FEERIC in ALICE}
From the results described above, it is however very difficult to anticipate the achievable charge gain
in the experimental working conditions. This is why it was foreseen to equip one of the 72 RPCs in the ALICE 
cavern with FEERIC electronics with final ASIC, in order to evaluate the performance during LHC Run 2.
In February 2015, 39 cards were installed on one RPC with strips of 2 and 4 cm,
placed in the fourth plane of the muon trigger.
The performance in terms of efficiency, cluster size, current, single counting rate, stability and robustness 
are now discussed exploiting the 2015 data taking in pp collisions at $\sqrt{s}=5$ and 13 TeV
and in Pb--Pb collisions at $\sqrt{s_{\mathrm{NN}}}=5$ TeV.

\subsection{Working point determination}
\begin{figure}[htbp]
 \centering
 \includegraphics[width=0.496\textwidth]{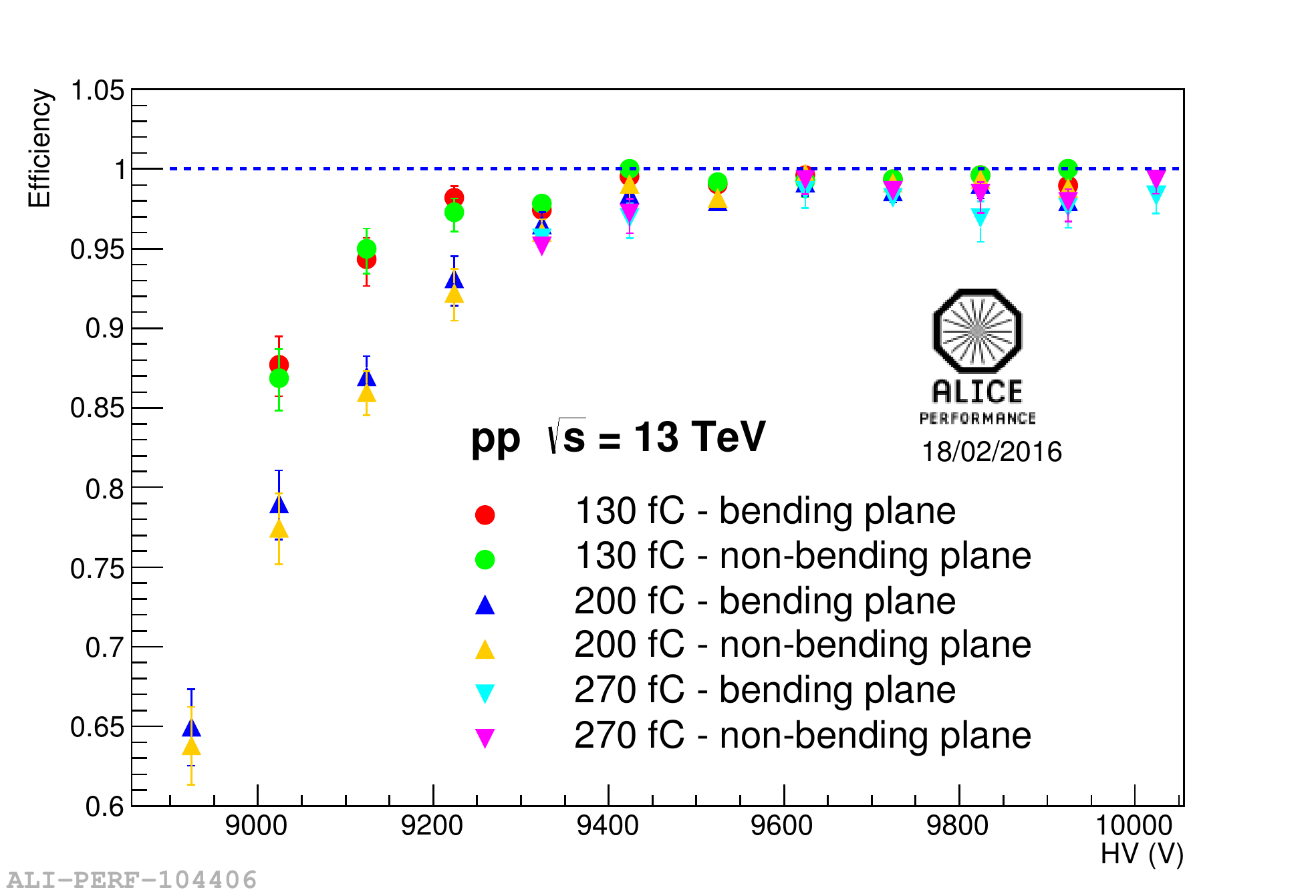}
 \includegraphics[width=0.496\textwidth]{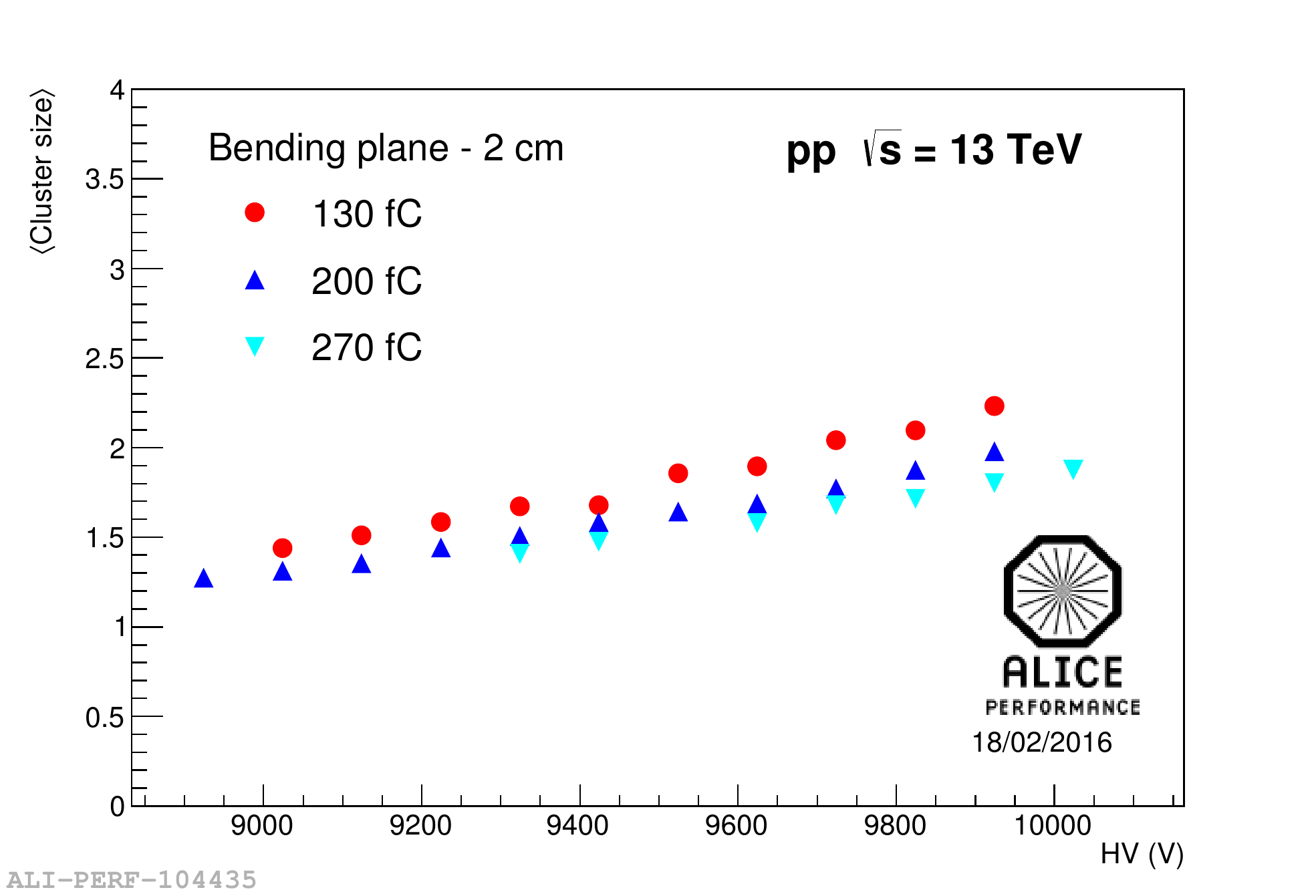}
 \caption{Efficiency plateau (left) and cluster size measurements for strips of 2 cm (right) in 
 pp collisions at 13 TeV. Colors refer to different discrimination thresholds.}
 \label{fig: working_point}
\end{figure}

The working point of the RPC equipped with FEERIC cards has been determined by means 
of a HV scan performed in June 2015 during low luminosity pp runs at 13 TeV, for three thresholds.
The results are shown in figure \ref{fig: working_point} (left).
Full efficiency is reached for all threshold values and the results are
very similar for positive (bending plane) and negative (non-bending plane) polarities.

The HV dependence of the cluster size for strips of 2 cm is also shown in figure \ref{fig: working_point} 
(right). Measurements are larger for the lower threshold values and increase with HV.
As already noticed during the tests with cosmic rays, in maxi-avalanche mode (with ADULT
cards with a threshold of 7 mV) the cluster size is lower, $\sim$1.4 for strips of 2 cm width 
\cite{RPC_run1}.

By considering these results and the fact that no noise is observed in ALICE cavern
for a threshold voltage above 33 mV ($\sim$60 fC) for all FEERIC card formats and polarities,
the working point has been set at 9375 V for a 70 mV (130 fC) threshold.
For comparison, the working  point of this RPC when equipped with ADULT cards was 10125 V for a threshold of 7 mV 
(maxi-avalanche mode without FEE amplification). The HV difference is of 750 V, in line with
what was observed during preliminary tests with cosmic rays.

\subsection{Efficiency stability}
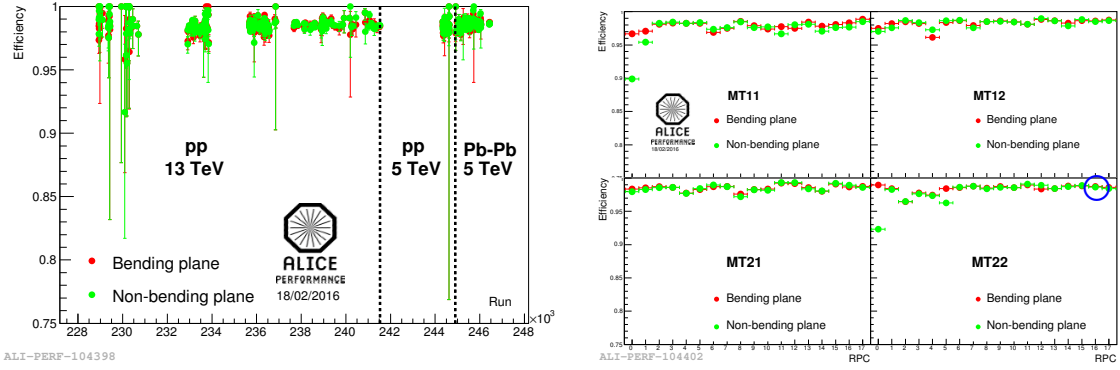
\begin{figure}[htbp]
 \centering
 \includegraphics[width=0.51\textwidth]{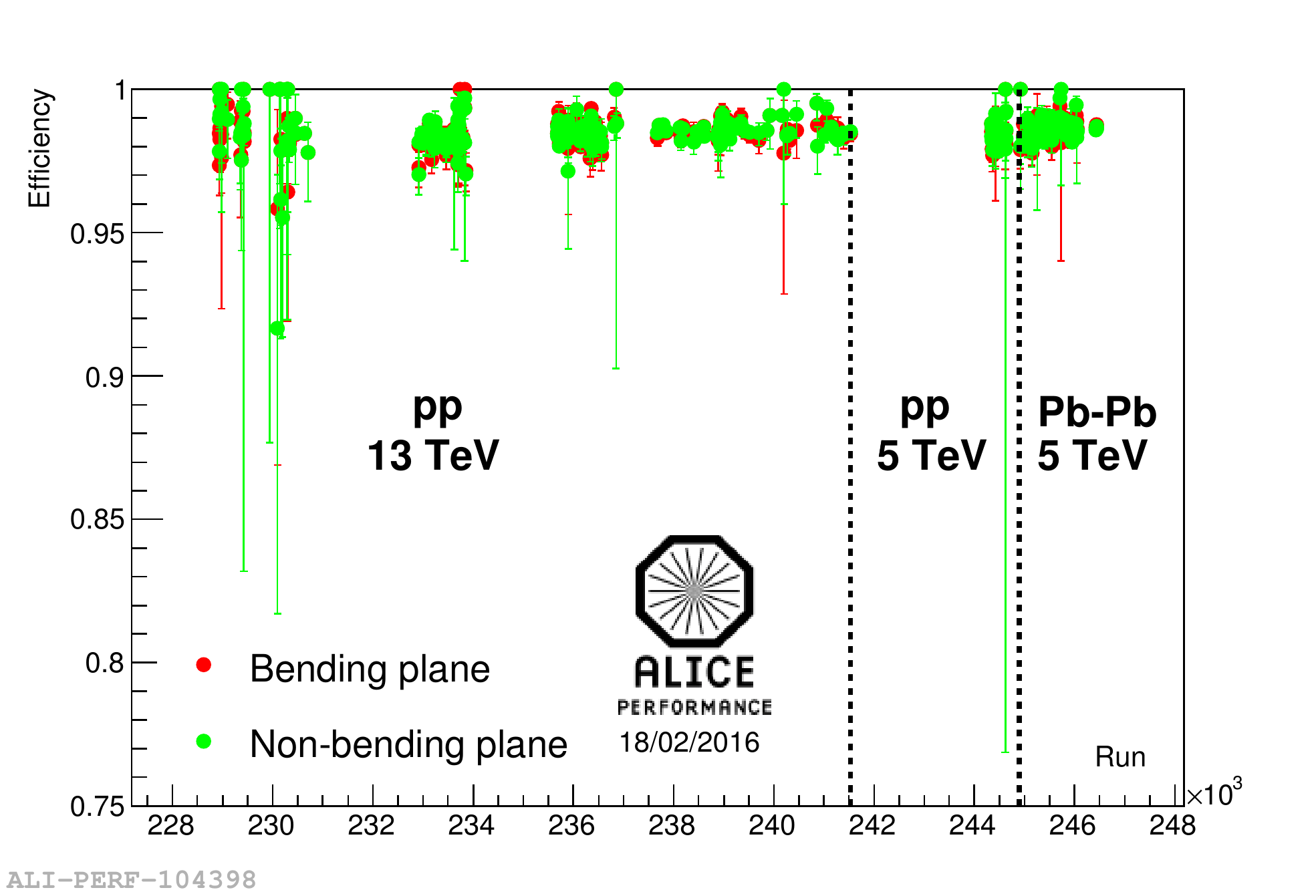}
 \includegraphics[width=0.475\textwidth]{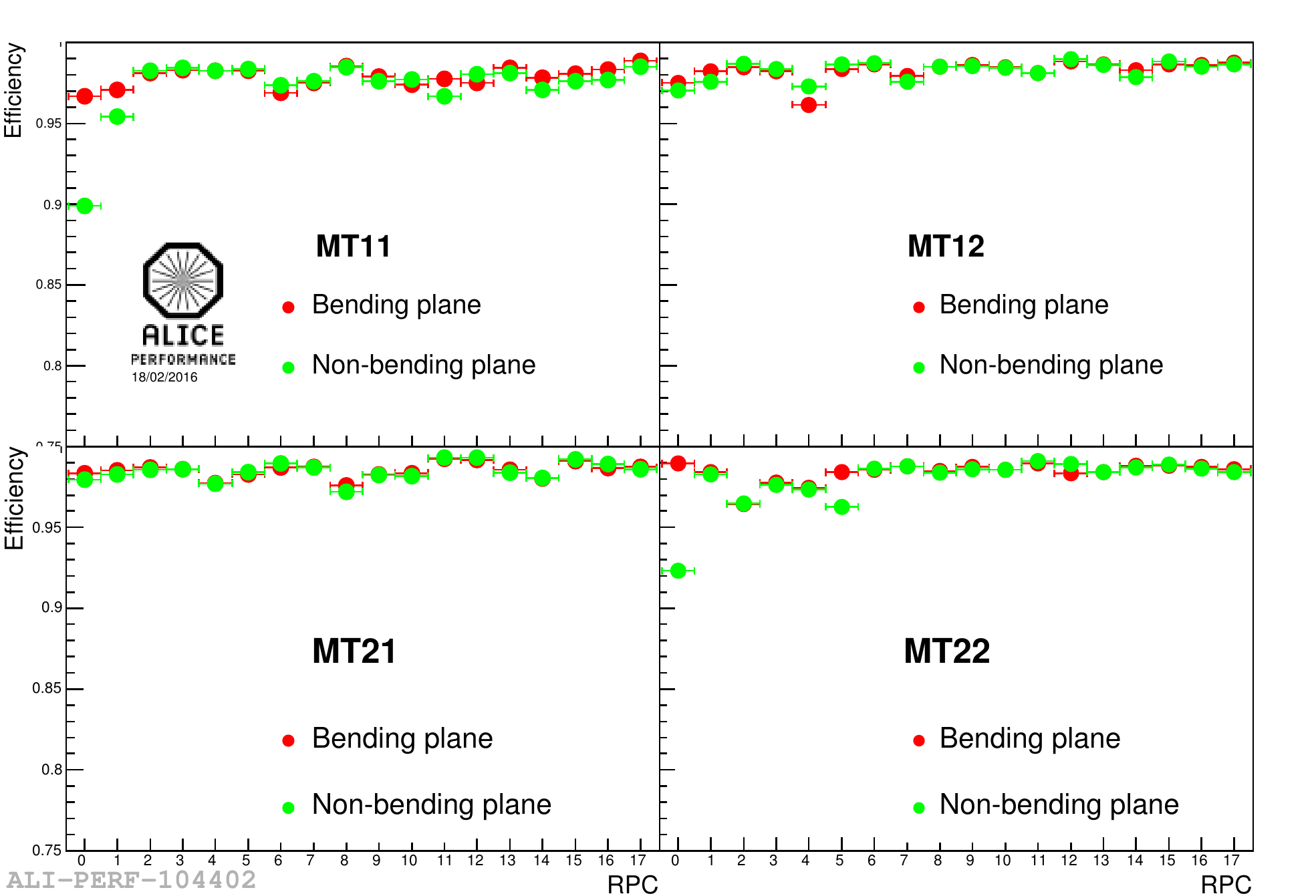}
 \caption{Left: efficiency of the RPC equipped with FEERIC measured in 2015. Right: average efficiency
 of the 72 RPCs composing the ALICE muon trigger. The blue circle indicates the chamber equipped with 
 the new electronics. Red (green) points indicate the bending (non-bending) plane.}
 \label{fig: efficiency}
 
\setlength{\unitlength}{0.01\textwidth}
\begin{picture}(0,0)\sffamily
\thicklines
 \put(45.7,30.4){\textcolor{blue}{\circle{2}}}
\end{picture}
\end{figure}

Figure \ref{fig: efficiency} (left) shows the time dependence of the efficiency of the RPC equipped
with FEERIC. Measurements demonstrate a very high efficiency ($\sim$97\%) for both bending and non-bending
planes. The stability is also evident for the three colliding systems characterizing the 2015
data taking.

The comparison with the other chambers is shown in figure \ref{fig: efficiency} (right). It displays
the efficiency of the 71 RPCs still equipped with the non-amplified
electronics and the chamber with the new front-end in the blue circle.
The values are averaged over the whole 2015. The efficiency of the chamber with FEERIC is perfectly in agreement
with the results obtained for the other RPCs or slightly better.

\subsection{Dark current and dark rate time dependence}
\begin{figure}[htbp]
 \centering
 \includegraphics[width=0.54\textwidth]{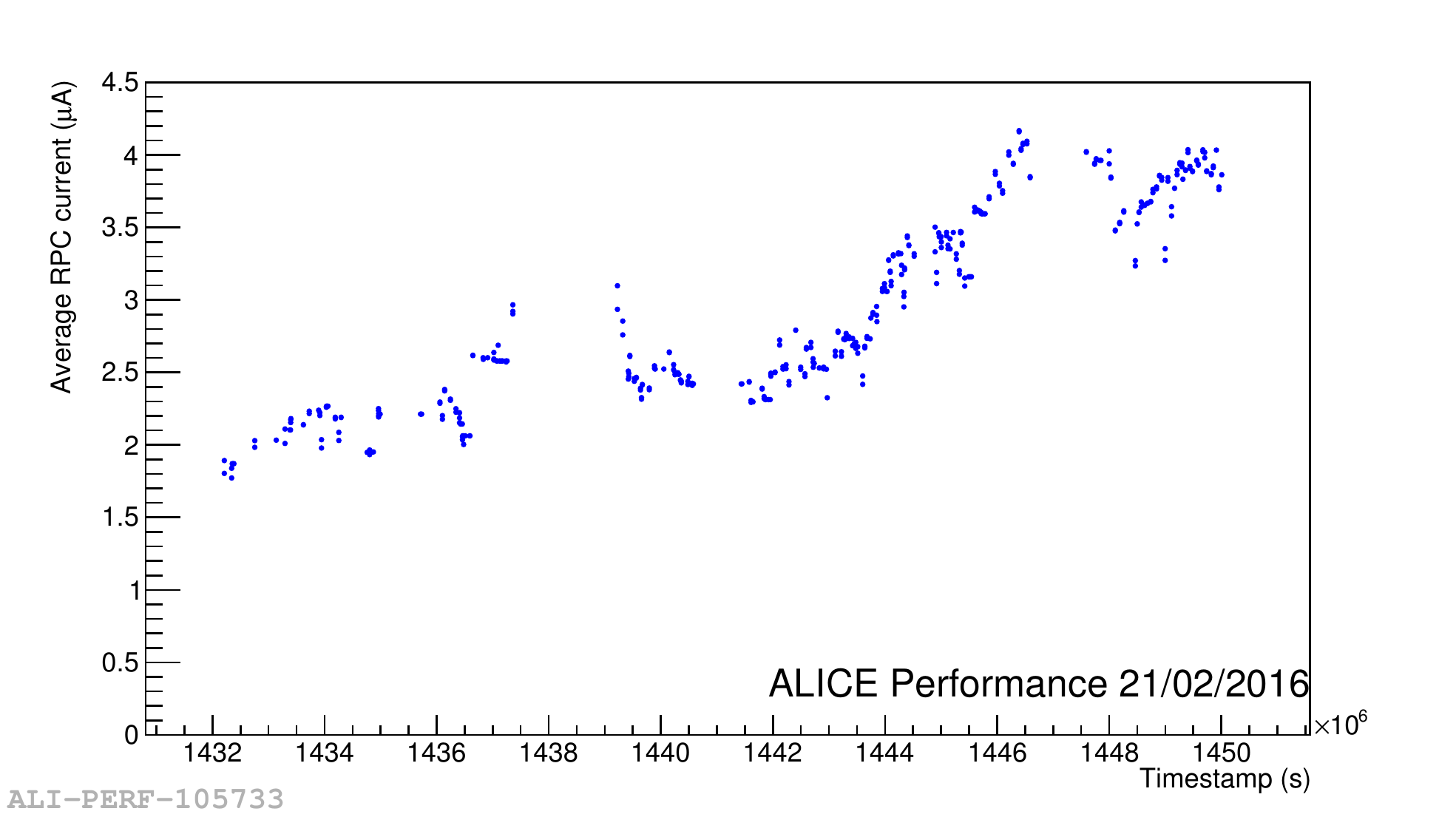}
 \includegraphics[width=0.45\textwidth]{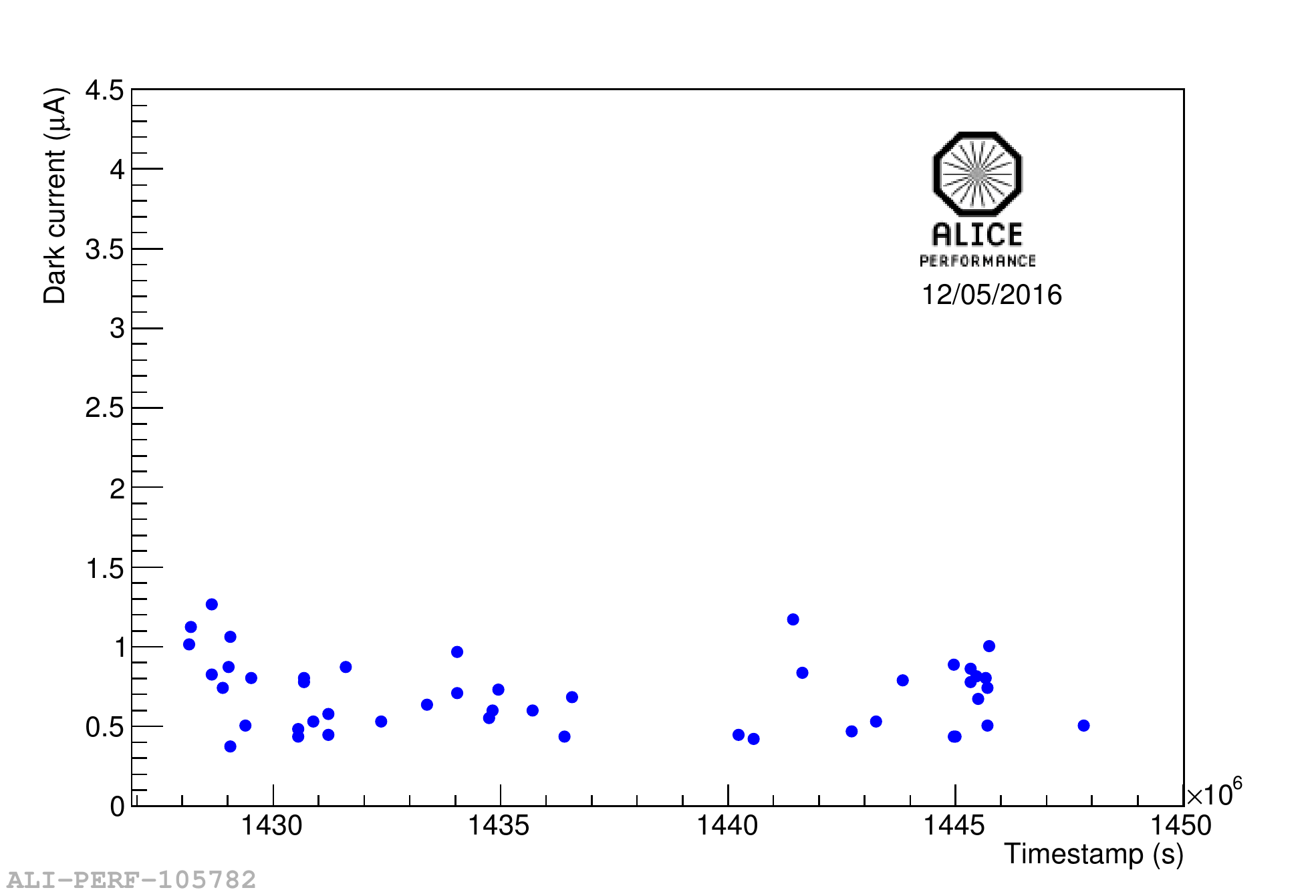}
 \caption{Average dark current drawn by the 71 RPCs equipped with ADULT (left) compared
 to the dark current of the RPC with FEERIC (right).}
 \label{fig: dark_current}
\end{figure}

Dark currents have been measured with dedicated runs in the absence of collisions, with the HV at their nominal values.
The average dark current of the 71 RPCs equipped with ADULT at nominal voltage ranges from 2 $\mu$A to
4 $\mu$A, increasing with time as can be seen in figure \ref{fig: dark_current} (left).
Conversely, the dark current of RPC with FEERIC is much lower ($\sim$0.6 $\mu$A) and stable throughout
the data taking, as shown in the right part.

\begin{figure}[htbp]
 \centering
 \includegraphics[width=0.54\textwidth]{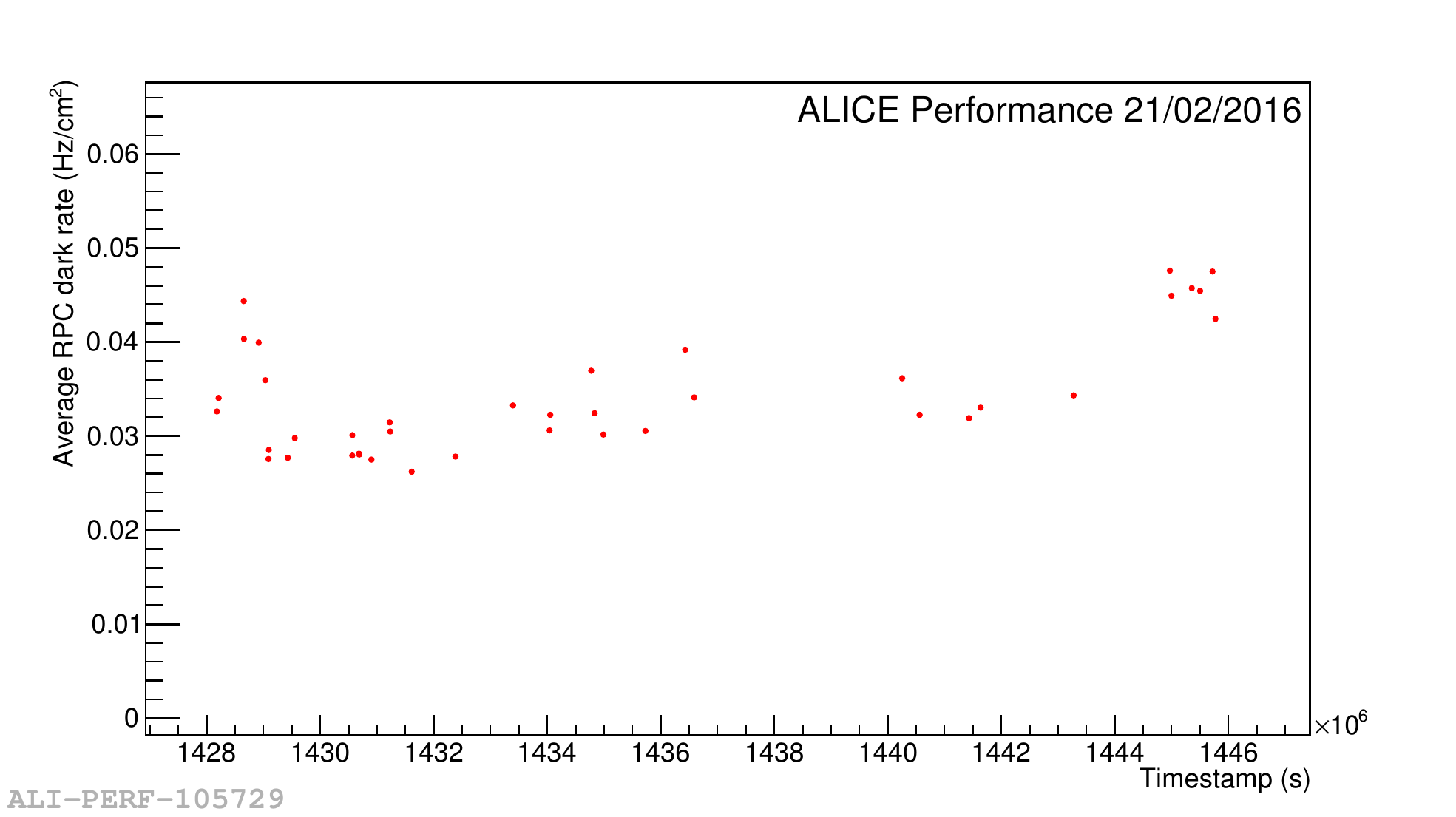}
 \includegraphics[width=0.45\textwidth]{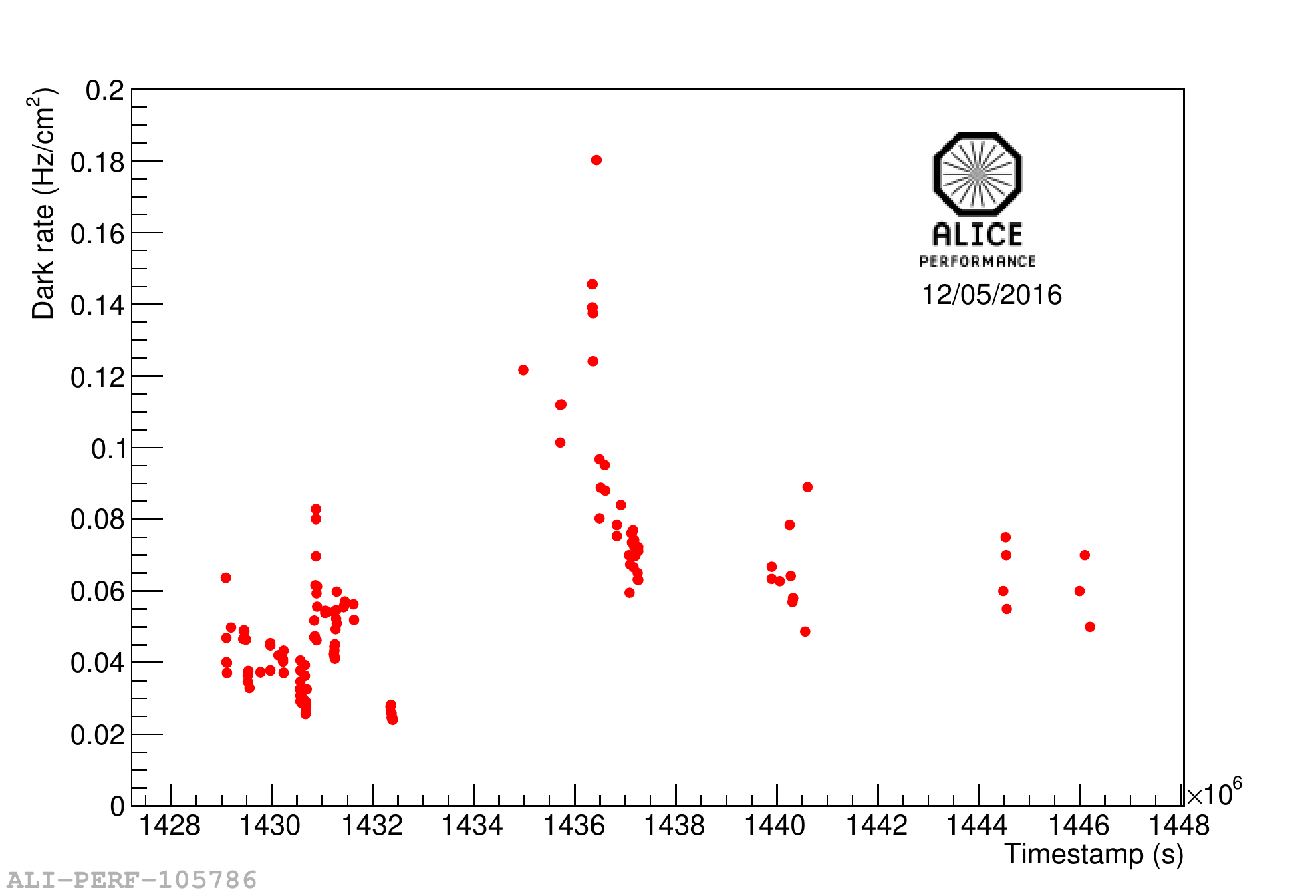}
 \caption{Average dark rate of the RPCs with ADULT (left) compared
 to the dark rate of the RPC with FEERIC (right).}
 \label{fig: dark_rate}
\end{figure}

The dark rate has also been measured.
This measurement is obtained from the strip scalers and hence it is dead time free. The average
dark rate of RPCs with ADULT is below 0.05 Hz/cm$^2$ (figure \ref{fig: dark_rate}, left)
and stable in time, while the dark rate of RPC with FEERIC is larger, but usually
below 0.1 Hz/cm$^2$.

\subsection{Evaluation of the charge reduction with FEERIC}

\begin{figure}[htbp]
 \centering
 \includegraphics[width=0.7\textwidth]{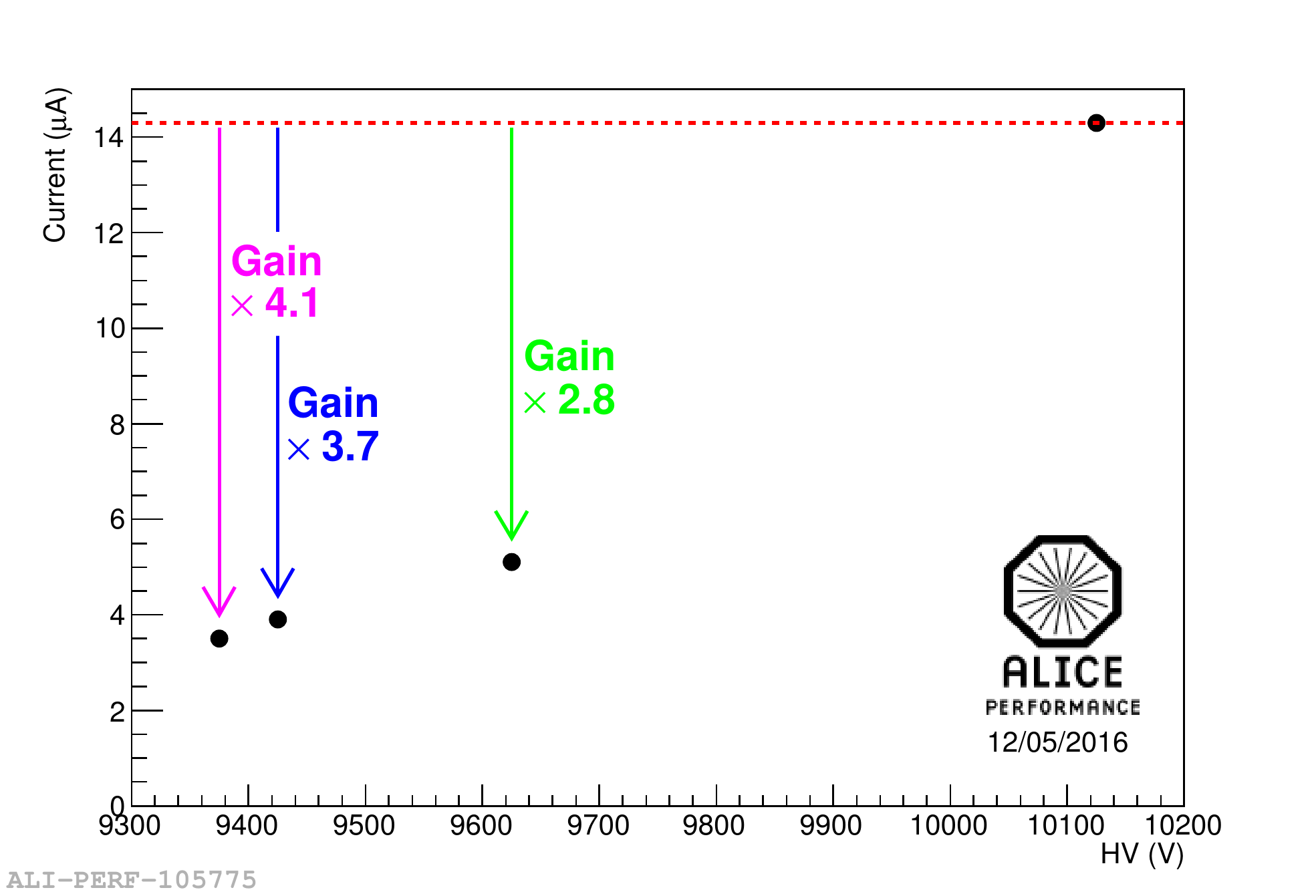}
 \caption{Current drawn in pp collisions by the RPC equipped with FEERIC at different HV.
 The gain with respect to the old working point is also shown.}
 \label{fig: current}
\end{figure}

As previously mentioned, the main goal of this front-end electronics upgrade is to reduce
the total charge produced in the gas gap, in order to prevent aging effects and to improve the rate capability.
In order to evaluate the gain in charge obtained in the new working conditions, the RPC current has been measured
at fixed interaction rate of $\sim$400 kHz in July 2015 after subtraction of the dark component. Measurements have been
performed at the HV value of 9375 V, corresponding to the working point with FEERIC, and at the HV value
of 10125 V (ADULT working point). The results are shown in figure \ref{fig: current}, together with two
intermediate HV measurements. The gain in current (and therefore in charge) obtained is more than a factor of 4.

\section{Conclusions and perspectives}

In summary, a completely new front-end electronics has been designed with the aim to increase the
rate capability and limit aging effects.
The performance of one RPC equipped with the new cards have been studied in a
dedicated cosmic data taking and in pp and Pb--Pb collisions.
It has been possible to lower the working point by 750 V maintaining an efficiency larger than 95\%.
The main goal of the FEERIC upgrade project has been achieved: a gain in current (and therefore in total charge)
of a factor 4 has been reached as compared to the present operations in maxi-avalanche.
Performance related to cluster size and single counting rate are also within specifications. This study confirmed the
robustness and the reliability on a long time scale of FEERIC in realistic conditions.

\end{document}